\documentclass[aps,prl,showpacs,twocolumn,superscriptaddress,10pt]{revtex4-1}

\usepackage{graphicx}
\usepackage{textcomp} 
\usepackage[usenames]{color}
\usepackage{epsfig}
\usepackage{verbatim}
\usepackage{listings}
\usepackage{amsmath}
\usepackage{fancyhdr}

\begin{document}

\newcommand{\mup}{\ensuremath{\mu {\mathnormal{p}}}}
\newcommand{\mud}{\ensuremath{\mu {\mathnormal{d}}}}

\newcommand{\ppmu}{\ensuremath{[(p p \mu)^+]^*}}
\newcommand{\ddmu}{\ensuremath{[(d d \mu)^+]^*}}
\newcommand{\DDmu}{\ensuremath{(d d \mu)^+}}
\newcommand{\DTmu}{\ensuremath{(d t \mu)^+}}

\newcommand{\pmue}{\ensuremath{[(p \mu e)^+]^*}}

\newcommand{\ppmupee}{\ppmu{} \ensuremath{p e e}}
\newcommand{\ddmudee}{\ddmu{} \ensuremath{d e e}}

\newcommand{\mumi}{\ensuremath{\mu^{-}}}
\newcommand{\Ekin}{\ensuremath{\mathrm E_{\mathrm{kin}}}}

\newcommand{\OneS}{{\ensuremath{1S}}}
\newcommand{\TwoS}{{\ensuremath{2S}}}
\newcommand{\TwoP}{{\ensuremath{2P}}}

\newcommand{\LyA}{Lyman-$\alpha$}

\newcommand{\Ka}{\ensuremath{\mathrm K_{\alpha}}}
\newcommand{\Kb}{\ensuremath{\mathrm K_{\beta}}}
\newcommand{\Kr}{\ensuremath{\mathrm K_{{\geq} \gamma}}}

\newcommand{\chisq}{\ensuremath{\chi^{2}}}
\newcommand{\mus}{\ensuremath{\mu \mathrm{s}}}

\newcommand{\muN}{\ensuremath{\mu {\mathrm{N}}}}
\newcommand{\muO}{\ensuremath{\mu {\mathrm{O}}}}
\newcommand{\muC}{\ensuremath{\mu {\mathrm{C}}}}
\newcommand{\Htwo}{\ensuremath{\mathrm{H_{2}}}}
\newcommand{\Dtwo}{\ensuremath{\mathrm{D_{2}}}}

\newcommand{\Otwo}{\ensuremath{\mathrm{O_{2}}}}
\newcommand{\Ntwo}{\ensuremath{\mathrm{N_{2}}}}

\newcommand{\dele}{\emph{del-e}}
\newcommand{\xdele}{\emph{x-ray {+} del-e}}

% cascade times
\newcommand{\tcasc}{\ensuremath{\tau_{\mathrm{casc}}}}
\newcommand{\tcascH}{\ensuremath{\tcasc^{\mup}}}
\newcommand{\tcascD}{\ensuremath{\tcasc^{\mud}}}

\newcommand{\TcascKa}{\ensuremath{\tcasc^{\alpha}}}
\newcommand{\TcascKr}{\ensuremath{\tcasc^{\geq \gamma}}}
\newcommand{\TcascAvg}{\ensuremath{\tcasc^{\mathrm{avg}}}}

% time offsets
\newcommand{\dTcasc}{\ensuremath{\Delta t_{K}}}

% Yields
\newcommand{\Yka}{\ensuremath{Y_{K \alpha} / Y_{tot}}}
\newcommand{\Ykb}{\ensuremath{Y_{K \beta} / Y_{tot}}}
\newcommand{\Ykr}{\ensuremath{Y_{K {\geq} \gamma} / Y_{tot}}}

\newcommand{\Ykamup}{\ensuremath{0.803 \pm 0.012}}  

\newcommand{\Ykamud}{\ensuremath{0.78 \pm 0.02}}
\newcommand{\Ykbmud}{\ensuremath{0.06 \pm 0.01}}
\newcommand{\Ykrmud}{\ensuremath{0.16 \pm 0.03}}

% 2S population from yields
\newcommand{\PopTSmuH}{\ensuremath{2.76 \pm 0.17}}  % 3.09 +- 0.27 %
\newcommand{\PopTSmuD}{\ensuremath{3.1 \pm 0.3}}  % 3.09 +- 0.27 %

%-----------------------
% analysis results:

%chi^2/dof
\newcommand{\Chimup}{252.5}
\newcommand{\Chimud}{258.5}
\newcommand{\Dofmup}{247}
\newcommand{\Dofmud}{256}

\newcommand{\Tauavmup}{\ensuremath{25.3\pm1.4}}
\newcommand{\Taukamup}{\ensuremath{25.0\pm1.5}}
\newcommand{\Taukrmup}{\ensuremath{27.7\pm3.0}}

\newcommand{\Tauavmud}{\ensuremath{31.4\pm1.8}}
\newcommand{\Taukamud}{\ensuremath{31.6\pm1.7}}
\newcommand{\Taukrmud}{\ensuremath{28.6\pm1.9}}

\newcommand{\SHORT}{\ensuremath{\mathrm{short}}}
\newcommand{\LONG}{\ensuremath{\mathrm{long}}}
\newcommand{\TOTAL}{\ensuremath{\mathrm{total}}}

% fit parameters
\newcommand{\TFit}{\ensuremath{\tilde{\tau}}}

\newcommand{\AfitShort}{\ensuremath{\tilde{\varepsilon}_\SHORT}}
\newcommand{\TfitShort}{\ensuremath{\tilde{\tau}_\SHORT}}
\newcommand{\AfitLong}{\ensuremath{\tilde{\varepsilon}_\LONG}}
\newcommand{\TfitLong}{\ensuremath{\tilde{\tau}_\LONG}}

% FITted lifetimes, amplitudes
% Results.ods 7.6.2013, 15:36
\newcommand{\Tsmup}{\ensuremath{104\,^{+15}_{-13}}}      % ns
\newcommand{\Tsmud}{\ensuremath{142\,^{+29}_{-33}}}      % ns
\newcommand{\Tlmud}{\ensuremath{1.09\,^{+0.70}_{-0.49}}}  % mus

\newcommand{\Asmup}{\ensuremath{1.32\,^{+0.42}_{-0.31}}}  % %
\newcommand{\Asmud}{\ensuremath{1.02\,^{+0.41}_{-0.23}}}  % %
\newcommand{\Almud}{\ensuremath{0.11\,^{+0.07}_{-0.03}}}  % %
\newcommand{\AlmudY}{\ensuremath{0.085\,^{+0.058}_{-0.026}}}  % % \Almud * yield (0.8)

% MC corrections 
% Results.ods 7.6.2013, 15:36
\newcommand{\CTsmup}{\ensuremath{0.96\,^{+0.06}_{-0.03}}}
\newcommand{\CAsmup}{\ensuremath{1.56\,^{+0.10}_{-0.05}}}

\newcommand{\CTsmud}{\ensuremath{0.97\,^{+0.11}_{-0.06}}}  % errors *2
\newcommand{\CAsmud}{\ensuremath{1.59\,^{+0.20}_{-0.13}}}  % errors *2
\newcommand{\CTlmud}{\ensuremath{1.06 \pm 0.11}}        % errors *2

% physics results
% Results.ods 10.7.
\newcommand{\AmpTwoSmup}{\ensuremath{1.73\,^{+0.56}_{-0.42}}}  % Yields from FK
\newcommand{\AmpTwoSmud}{\ensuremath{1.35\,^{+0.57}_{-0.33}}}  % our new muD yields

\newcommand{\TauTwoSmup}{\ensuremath{100\,^{+16}_{-13}}}
\newcommand{\TauTwoSmud}{\ensuremath{138\,^{+32}_{-34}}}

\newcommand{\TauTwoSlmud}{\ensuremath{1.15\,^{+0.75}_{-0.53}}}
\newcommand{\AmpTwoSlmud}{\ensuremath{0.17\,^{+0.15}_{-0.09}}} 

% mud: Eps_Total - Eps_Short = expected Eps_Long
\newcommand{\AmpLongExpected}{\ensuremath{1.7\,^{+0.4}_{-0.7}}} % muD

% physical parameters
\newcommand{\EpsShort}{\ensuremath{\varepsilon^\SHORT_\TwoS}}
\newcommand{\TauShort}{\ensuremath{\tau^\SHORT_\TwoS}}

\newcommand{\EpsLong}{\ensuremath{\varepsilon^\LONG_\TwoS}}
\newcommand{\TauLong}{\ensuremath{\tau^\LONG_\TwoS}}
\newcommand{\TauMu}{\ensuremath{\tau_\mu}}

\newcommand{\EpsTotal}{\ensuremath{\varepsilon^\TOTAL_\TwoS}}

\newcommand{\CORR}{\ensuremath{C}}
\newcommand{\CorrEpsLong}{\ensuremath{\CORR^\varepsilon_\LONG}}
\newcommand{\CorrEpsShort}{\ensuremath{\CORR^\varepsilon_\SHORT}}

\newcommand{\CorrTauLong}{\ensuremath{\CORR^\tau_\LONG}}
\newcommand{\CorrTauShort}{\ensuremath{\CORR^\tau_\SHORT}}

\hyphenation{quench-ing}

\newcommand\CHECK[1]{{\color{red} #1}}

%Title of paper
\title{Lifetime and population of the {\boldmath $2S$} state in 
  muonic hydrogen and deuterium}

\def\AVEIRO{I3N, Departamento de F\'isica, Universidade de Aveiro, 3810--193 Aveiro, Portugal.} 
\def\COIMBRA{Departamento de F\'isica, Universidade de Coimbra, 3004--516 Coimbra, Portugal.}
\def\DAUSINGER{Dausinger \& Giesen GmbH, Roteb\"uhlstr.~87, 70178 Stuttgart, Germany.}
\def\IPPZURICH{Institute for Particle Physics, ETH Zurich, 8093 Zurich, Switzerland.} 
\def\LKB{Laboratoire Kastler Brossel, \'Ecole Normale Sup\'erieure, CNRS, and 
  \penalty -100  Universit\'e P.~et M.~Curie -- Paris 6, Case 74; 
  \penalty -100 4, place Jussieu, 75252 Paris, CEDEX 05, France.}
\def\MPQ{Max--Planck--Institut f{\"u}r Quantenoptik, 85748 Garching, Germany.} 
\def\PRINCETON{Department of Chemistry, Princeton University, Princeton, NJ08544--1009, USA.}
\def\PSI{Paul Scherrer Institute, 5232 Villigen--PSI, Switzerland.}
\def\STUTTGARD{Institut f{\"u}r Strahlwerkzeuge, Universit{\"a}t Stuttgart, 70569 Stuttgart, Germany.}
\def\TSINGHUA{Physics Department, National Tsing Hua University, Hsinchu 300, Taiwan.}
\def\UNIFR{D\'epartement de Physique, Universit\'e de Fribourg, 1700 Fribourg, Switzerland.} 
\def\YALE{Physics Department, Yale University, New Haven, CT 06520--8121, USA.}

\author{Marc~Diepold}
\affiliation{\MPQ}

\author{Fernando~D.~Amaro}
\affiliation{\COIMBRA} 

\author{Aldo Antognini}
\affiliation{\MPQ}
\affiliation{\IPPZURICH}

\author{Fran\c{c}ois~Biraben}
\affiliation{\LKB} 

\author{Jo\~{a}o~M.~R.~Cardoso}
\affiliation{\COIMBRA}  
 
\author{Daniel~S.~Covita}
\affiliation{\AVEIRO} 

\author{Andreas~Dax}
\affiliation{\YALE}  
 
\author{Satish~Dhawan}
\affiliation{\YALE}  

\author{Luis~M.~P.~Fernandes}
\affiliation{\COIMBRA}  

\author{Adolf~Giesen}
\affiliation{\STUTTGARD} 
\affiliation{\DAUSINGER} 

\author{Andrea~L.~Gouvea}
\affiliation{\COIMBRA}  

\author{Thomas~Graf}
\affiliation{\STUTTGARD} 

\author{Theodor~W.~H{\"a}nsch}
\thanks{also at Ludwig-Maximilians-Universit\"at, Munich, Germany.}
\affiliation{\MPQ}

\author{Paul~Indelicato}
\affiliation{\LKB}

\author{Lucile~Julien}
\affiliation{\LKB}

\author{Cheng-Yang Kao}
\affiliation{\TSINGHUA}

\author{Paul~Knowles}
\affiliation{\UNIFR}

\author{Franz~Kottmann}
\affiliation{\IPPZURICH}

\author{Eric-Olivier~Le~Bigot}
\affiliation{\LKB}

\author{Yi-Wei~Liu}
\affiliation{\TSINGHUA}

\author{Jos\'{e}~A.~M.~Lopes}
\thanks{also at Instituto Polit\'ecnico de Coimbra, ISEC, 3030--199, Portugal.}
\affiliation{\COIMBRA} 

\author{Livia~Ludhova}
\affiliation{\UNIFR}

\author{Cristina~M.~B.~Monteiro}
\affiliation{\COIMBRA}  

\author{Fran\c{c}oise~Mulhauser}
\affiliation{\UNIFR}
\affiliation{\MPQ}

\author{Tobias~Nebel}
\affiliation{\MPQ}

\author{Fran\c{c}ois~Nez}
\affiliation{\LKB}

\author{Paul~Rabinowitz}
\affiliation{\PRINCETON} 

\author{Joaquim~M.~F.~dos~Santos}
\affiliation{\COIMBRA}  

\author{Lukas~A.~Schaller}
\affiliation{\UNIFR}

\author{Karsten~Schuhmann}
\affiliation{\IPPZURICH}
\affiliation{\DAUSINGER} 
\affiliation{\PSI}

\author{Catherine~Schwob}
\affiliation{\LKB}

\author{David~Taqqu}
\affiliation{\PSI}

\author{Jo\~{a}o~F.~C.~A.~Veloso}
\affiliation{\AVEIRO} 

\author{Jan Vogelsang}
\thanks{present address: Institut f{\"u}r Physik, Carl von Ossietzky Universit{\"a}t, 26129 Oldenburg, Germany.}
\affiliation{\MPQ}

\author{Randolf~Pohl}                        
\affiliation{\MPQ}

\collaboration{CREMA collaboration}
\noaffiliation

%\email[]{platzhalter}
\date{\today}

\bibliographystyle{mysty}

%===================================================================

\begin{abstract}
Radiative de-excitation (RD) of the 
metastable \TwoS{} state of muonic hydrogen and deuterium atoms has been observed. 
In muonic hydrogen, we improve the precision on lifetime and population
(formation probability) values
for the short-lived $\mup(\TwoS)$
component, and give an upper limit for RD of long-lived
$\mup(\TwoS)$ atoms.
In muonic deuterium at 1~hPa, \PopTSmuD\% of all stopped muons form 
\mud(\TwoS) atoms. The short-lived \TwoS{} component 
has a population of \AmpTwoSmud\%
and a lifetime of $\TauShort(\mud) = \TauTwoSmud$~ns. We 
see evidence for RD of 
long-lived $\mud(\TwoS)$ with a lifetime of $\TauLong(\mud) =
\TauTwoSlmud~\mus$.
This is interpreted as formation and decay of
excited muonic molecules.
\end{abstract}

%===================================================================

\pacs{36.10.Ee, 31.30.Gs, 33.50.Hv}

\maketitle

%===============================================

%\section{Introduction:}
\paragraph{{\bf Introduction} ---}

When negative muons \mumi{} are stopped in molecular  
hydrogen (H$_2$) or deuterium (D$_2$) gas, muonic hydrogen/deuterium atoms (\mup, \mud) are formed in highly
excited states with principal quantum number
$n \approx14$~\cite{LeonBethe:1962:mesonicH}.
Radiative and collisional de-excitation during the cascade leads to formation
of muonic hydrogen atoms in
the \OneS{} ground state or the \TwoS{} metastable 
state~\cite{Jensen:2002:Colldeexcite_Xsect,Jensen:2002:Colldeexcite_Cascade}.
The time interval between \mumi{} capture and its
arrival in the \OneS{} or \TwoS{} state
is given by the so-called cascade time \tcasc.
In muonic hydrogen,
$\tcascH$ was measured to be $ 37 \pm 5$~ns
at 0.6~hPa H$_2$ gas
pressure~\cite{Ludhova:2007:MupCascTime2S}.

The fraction of muons that actually reach the
\TwoS{} state can be calculated by the observed x-ray yields during 
the cascade~\cite{Anderhub:1977:Search2S,Egan:1981:LongLived2S,Anderhub:1984:KIntRatMup,Bregant:1998:IntRatiosMup}.
The fate of these atoms depends on their kinetic energy (\Ekin): In a
collision with a gas molecule, \TwoS{} atoms with $\Ekin >
0.3~\mathrm{eV}$~ can end up in the
short-lived \TwoP{} state which decays immediately to the ground state
via \LyA{} (\Ka) x-ray emission.
This 0.3~eV threshold energy in the laboratory frame 
corresponds to the center-of-mass \mup{} \TwoS{}-\TwoP{} Lamb shift splitting of 0.2~eV.
These fast de-excited \TwoS{} atoms constitute the so-called short-lived
\TwoS{} component with a population $\EpsShort = 1.70^{+0.80}_{-0.56}$\% of all created \mup{} atoms, 
and a lifetime of $\TauShort = 165^{+38}_{-29}$~ns in \mup{} at
0.6~hPa~\cite{Ludhova:2007:MupCascTime2S}.

In contrast, \TwoS{} atoms with \Ekin{} below the threshold constitute the
so-called long-lived \TwoS{} atoms.
In vacuum, the lifetime of the \mup(\TwoS) atoms is given by the
muon lifetime of $\TauMu{} \approx 2.2~\mus{}$, since the two-photon decay rate is negligibly small.
In gaseous environments, collisional processes provide additional
decay channels.
This was first observed in \mup, where a lifetime of $\TauLong{} \approx
1~\mus{}$ and a formation probability of 
$\EpsLong{} = 1.10\pm0.08\%$ \cite{Pohl:2001:PhD}
was measured for this long-lived component at 1~hPa~\cite{Pohl:2006:MupLL2S}.
Such a population was imperative for the measurement 
of the \TwoS{}-\TwoP{} Lamb shift in muonic hydrogen with laser spectroscopy~\cite{Pohl:2010:Nature,
Antognini:2013:Science}.

The dominant de-excitation mechanism of the long-lived \mup(\TwoS) atoms is via
non-radiative Coulomb de-excitation (CD) in a collision, 
leading to \mup(\OneS) atoms with a \Ekin{} of 900~eV~\cite{Pohl:2001:PhD,Pohl:2006:MupLL2S,Pohl:2001:MolecQuenchMup}.
% which is why earlier searches for long-lived \TwoS{} atoms had failed~\cite{}
Such behavior had been predicted as a result of molecular
effects~\cite{Froelich:1993:Sidepath} via resonant formation of
excited muonic molecular ions \ppmu{} and their subsequent decay.
This happens similar to the Vesman-mechanism~\cite{Vesman:1967:muCF} 
which is responsible for muon-catalyzed fusion via the formation of
molecular ions from the ground state.
Recently, the molecular origin of the observed CD has been
questioned, and ``direct'' CD in a \mup{} + H collision
has been proposed as the source of the 
observed \mup(\OneS) atoms with \Ekin{} of 900~eV~\cite{Korenman:coulombjetp,Popov:2011:FormQuench2s}.

Theoretical
studies~\cite{Lindroth:2003:PPMU_with_erratum,Kilic:2004:Decay_ppmu} have
predicted very different behavior for excited muonic deuterium molecular ions, 
\ddmu, for which no data existed. The dominant de-excitation channel of
the excited \ddmu{} ion should be radiative de-excitation (RD) with a
branching ratio (BR) around 70\%, 
compared to only about 2\% in the excited muonic hydrogen
molecular ion \ppmu.

Here we report on the measurement of the cascade time,
population, and lifetime of both the long-lived and the short-lived \TwoS{} state in
muonic deuterium, \mud, and we provide direct evidence for RD of
long-lived \mud(\TwoS) atoms.
Additionally we give more precise values of the cascade time,
population, and lifetime of the short-lived \TwoS{} state in \mup{} and provide an
upper limit for RD of long-lived \mup(\TwoS) atoms.

\paragraph{{\bf Experiment} ---}

The data presented here are acquired during the muonic Lamb shift
experiment~\cite{Pohl:2010:Nature,Antognini:2013:Science} in 2009.
For this experiment low energy muons (3--6~keV kinetic energy) were stopped in a 20~cm long
target filled with 1~hPa molecular gas (\Htwo{} or \Dtwo) at
20$^\circ$C\@. Twenty large-area avalanche photo diodes (APDs, $14 \times
14$~mm$^2$ active area each)
\cite{Fernandes:2003:NIM,Ludhova:2005:LAAPDs} placed above and below the target served to detect x-rays
between $\sim 1$~keV and 20~keV\@ that originated from the formed muonic atoms.
The 12 APDs with the best x-ray energy
resolution of $\sim 21\%$ (FWHM) at 1.9~keV, and time resolution of
25~ns were used for this analysis, in order to reduce the background as far as possible.

The measured x-ray energies arise from \Ka, \Kb, and \Kr{} lines in 
\mup{} (1.90, 2.25, and $\sim 2.45$~keV, respectively) and 
\mud{} (2.00, 2.36, and $\sim 2.58$~keV),
%CHANGE 1st round: Tobi 3.7
%
silver fluorescence lines near 3~keV,
a muonic carbon line (\muC; 4.75~keV), 
and various muonic nitrogen (\muN; 1.01, 1.67, 3.08, 6.65~keV) 
and muonic oxygen           (\muO; 1.32, 2.19, 4.02, 8.69~keV)
transitions.
The last two contributions originate from a small leak, creating a 0.55(5)\%
air admixture in the target gas, that had no influence on the actual Lamb
shift measurement.
A silver layer on the target cell walls emitted additional delayed
x-rays through fluorescence upon contact with muonic atoms. 

The primary background source of the experiment derives from muon decay
electrons that are detected in the APDs, or in four 
plastic scintillators, with a time resolution of $\sim10$~ns. 
The offline event selection used in the analysis requires that an eligible x-ray must be followed by the
detection of a ``delayed'' high-energy electron (\dele) in a time window $t_e -t_X \in [0.15
\dots 6]~\mus$ after the x-ray time (\xdele{} events).
Cuts on the electron
identification further minimize the electron-induced background in the x-ray spectra.

%===============================================
%===============================================
\paragraph{{\bf Data analysis and fit results} ---}

\begin{figure}[t]
\begin{center}
\includegraphics[angle=0,width=0.9\columnwidth]{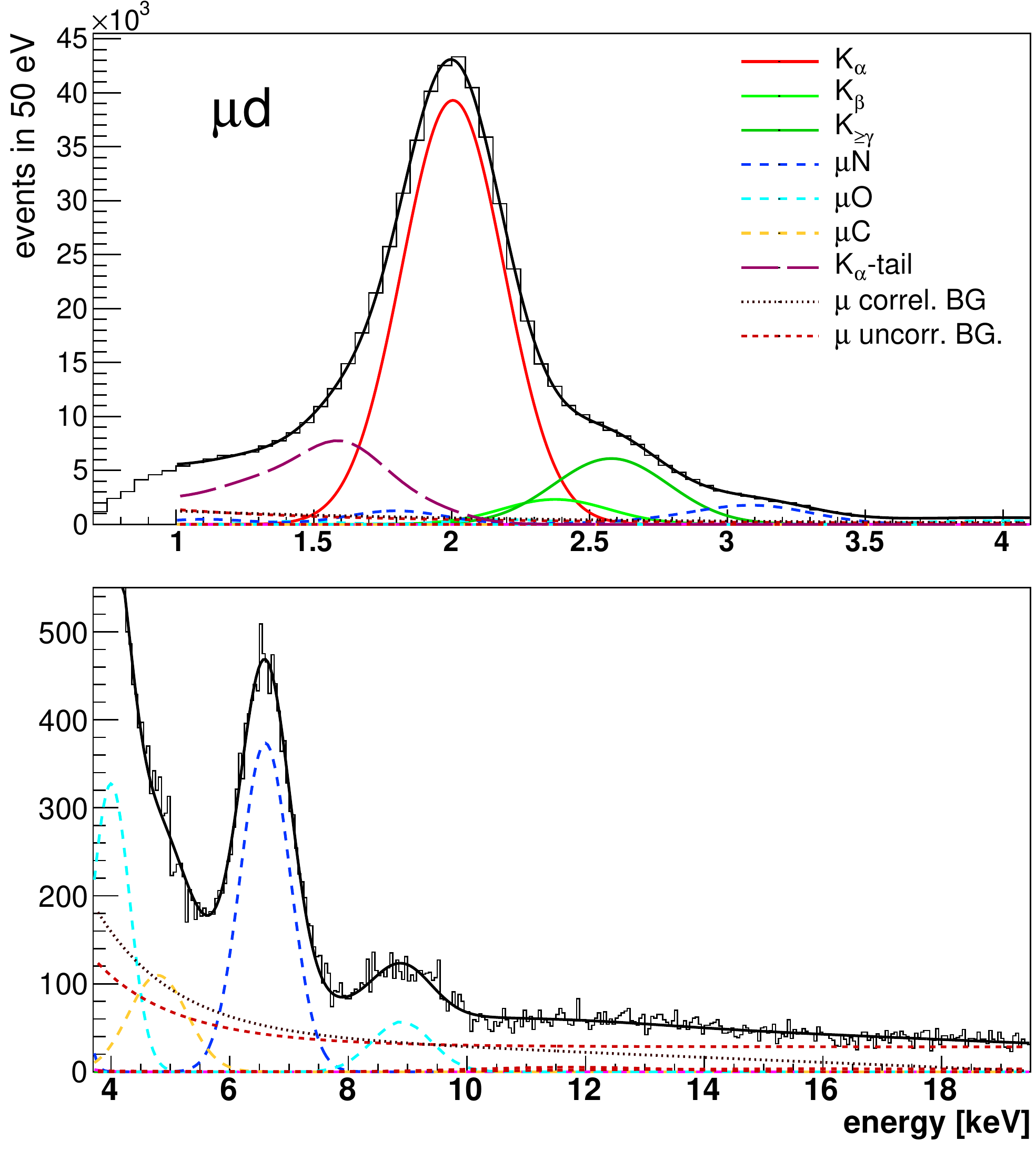}
\caption{ (Color online) Energy spectrum of all prompt x-rays
  in the muonic deuterium data. The plot is split to give a better overview over the
  contributions with low statistics.
}
\label{fig:fit_energy}
\end{center}
\vspace{-5mm}
\end{figure}

Following~\cite{Ludhova:2007:MupCascTime2S}, we first used time-integrated
  \emph{energy} spectra to determine the shape and position of each
of the various \mup/\mud, Ag, \muC, \muN, and \muO{} transitions,
as well as shapes of different background contributions.
% shapes were extracted from auxiliary spectra. 
Figure~\ref{fig:fit_energy} shows the \mud{} data.

%===============================================
\begin{figure*}[t]
  \begin{center}
    \setlength{\unitlength}{1\textwidth}
    \begin{picture}(1,0.50)(0,0)
      %\put(0,0) {\framebox(1,0.52){~}}
      \put(-0.005,0.245){
        \includegraphics[angle=0,width=1\textwidth]{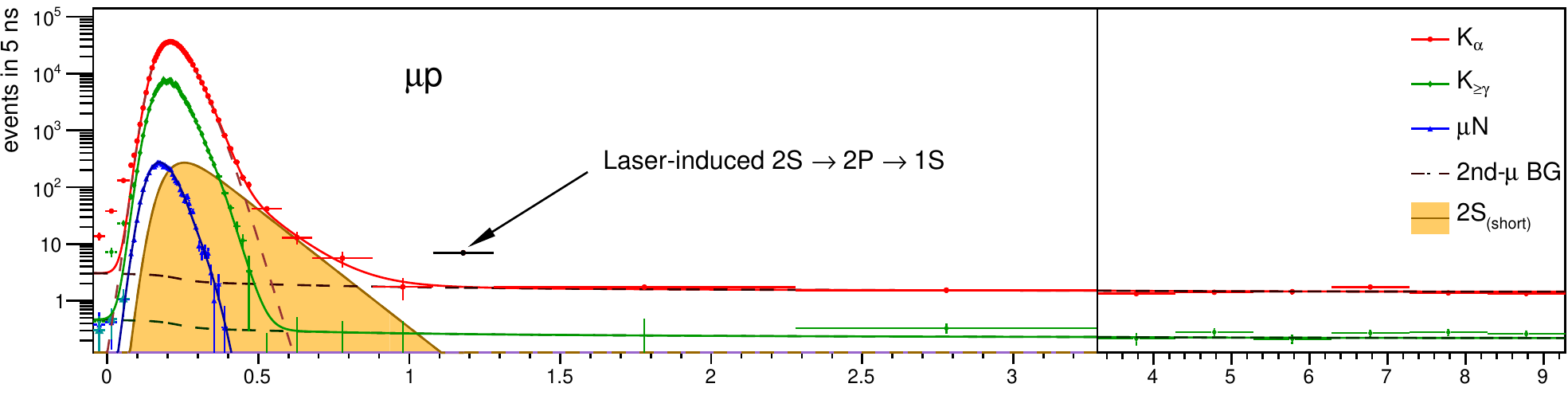}
      }
      \put(-0.005,0){
        \includegraphics[angle=0,width=1\textwidth]{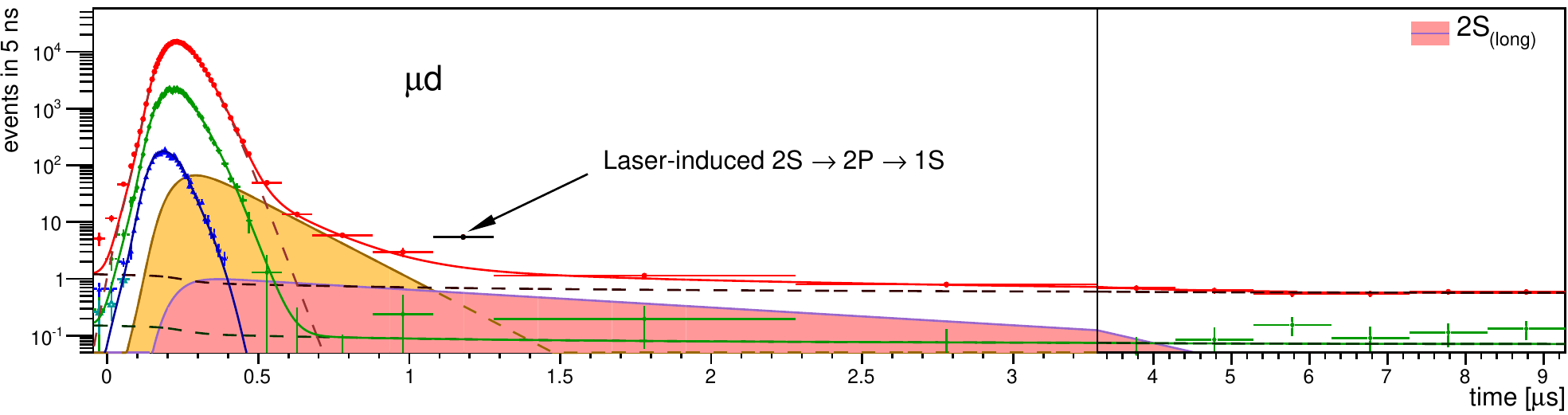}
      }
    \end{picture}
    \caption{(Color online) Time spectra and fits for muonic hydrogen (top) and
      deuterium (bottom). Note the break in the horizontal time axis.
      The stop time is given by the \muN{} and \muO{} (not shown) time spectra.
      The \Kr{} spectrum is the stop time, convoluted
      with an (exponential) cascade time distribution.
      The \Ka{} spectrum in \mup{} shows a short-lived exponential component
      from radiative quenching of fast \mup(\TwoS) atoms.
      In \mud, an additional  long-lived component is visible
      which we consider as evidence for
      radiative de-excitation of long-lived \mud(\TwoS) atoms after
      formation of excited molecular ions \ddmu.
      The \TwoS{} components are constructed by the convolution of the
      \Kr{} distribution with exponential lifetimes (see text).
      In \mup{} (\mud), 
      the prompt \Ka{} peak contains 881k (412k) events, and
      the fast component contains 11.6k (4.2k) events.
      The long-lived component in \mud{} contains 440 events.
      The fit region starts at 0.1~\mus{} to exclude early-time beam-correlated 
      background.
    }
    \label{fig:fit_time}
  \end{center}
  \vspace{-5mm}
\end{figure*}

%===============================================
In a second step,
we used the obtained parameterizations 
of the energy spectra to fit \emph{time slices} of the 
\xdele{} energy spectrum, varying only 
the amplitudes of x-ray lines and backgrounds.
The only exception to this is the position of the \Kr{} line
which was found to vary on the order of 50~eV during the time of the prompt
peak, as further discussed in the appendix.

The time-dependent x-ray amplitudes and their uncertainties gained by this
procedure provide the \emph{time spectra} of the
muonic hydrogen \Ka{} and \Kr, \muO, and \muN{} x-rays.
Sixty time slices of variable length (5~ns to 1~\mus, 
depending on the count rate)
span the time up to 9~\mus{} after muon entry.
The obtained time spectra are shown in Fig.~\ref{fig:fit_time}.

To obtain the final results, the
\muN, \muO, and muonic hydrogen
\Ka{} and \Kr{} time spectra
were fitted simultaneously (Fig.~\ref{fig:fit_time}).
The \Kb{} spectrum was found to behave as expected but was not further
considered here, because it strongly overlaps
with both the \Ka{} and \Kr{} contributions.
The fit function used connects different components on the
basis of prior knowledge of the muonic atom cascade:

\emph{Muon stop time}:
The muon stop time is given by the \muN{} and \muO{} time spectra, owing to
the negligibly short cascade time $\sim 10^{-10}$~s for these high-$Z$
atoms~\cite{Bracci:1978:CoulDeexMuH}. 
The same phenomenological shape was found to parameterize both the \muN{} and \muO{} time
spectra with a good $\chi^2$.

\emph{Muonic hydrogen/deuterium \Kr}:
The \Kr{} time spectrum is obtained by convoluting the muon stop time 
with the \Kr{} cascade time distribution. This distribution
is the sum of an exponential with the time constant \TcascKr{}, and a sharp peak at $t=0$,
as was suggested by cascade simulations~\cite{Jensen:2002:Colldeexcite_Xsect,Jensen:2002:Colldeexcite_Cascade}.
We find $\TcascKr{} = \Taukrmup$~ns and $\Taukrmud$~ns for \mup{} and \mud,
respectively.

Uncorrelated muons stopping in the target at random times give rise to a ``2nd-$\mu$'' background in the \Kr{} time spectrum which is essentially flat at
times after the prompt peak. The exact shape has been deduced from the data and shows
the expected decrease of event acceptance as a function of time.

\emph{Muonic hydrogen/deuterium \Ka}:
Prompt peak and 2nd-$\mu$ background of the \Ka{} time spectrum are fitted
similarly to the \Kr{} spectrum described above, with the only exception that a purely 
exponential cascade time distribution is used (no peak at $t=0$).
We find a \Ka{} cascade time $\TcascKa =
\Taukamup$~ns and $\Taukamud$~ns in \mup{} and \mud{} respectively.
These values agree with the respective \Kr{} cascade times, so we 
forced them to be equal for the final analysis 
(\TcascAvg{} in Table~\ref{tab:amp_tau}).

In addition, \Ka{} x-rays can originate from fast or slow RD
of the metastable \TwoS{} state. The \TwoS{} state is populated from the
same $P$ states that produce the \Kr{}
x-rays~\cite{Anderhub:1977:Search2S}, and is modeled to decay 
(bi-)exponentially. 
Hence, we use a convolution of the \Kr{} time spectrum with
exponentials to parameterize the \TwoS{} signal.
Short- and long-lived \TwoS{} atoms are modeled to give rise to a 
fast $(O(100~{\mathrm{ns}}))$ and slow $(O(1~\mus))$ decay time, respectively.

The time spectrum of \emph{muonic hydrogen} is fitted very well assuming 
a single fast exponential from the radiatively quenched short-lived \TwoS{} states
(Fig.~\ref{fig:fit_time}), with a \chisq/DOF = \Chimup/\Dofmup{}
for the simultaneous fits of all time spectra.
The \Ka{} time spectrum alone has a \chisq/DOF of 
48.8/48.
The fitted amplitude \AfitShort(\mup) = \Asmup\% 
and lifetime         \TfitShort(\mup) = \Tsmup~ns
are corrected as described below to obtain the physical values in
Table~\ref{tab:amp_tau}. 

For \emph{muonic deuterium}, both short- and long-lived components
have to be included in
the \Ka{} fit function to achieve a \chisq{}/DOF = \Chimud{}/\Dofmud.
The short-lived component has a fitted
amplitude      \AfitShort(\mud) = \Asmud\%
and lifetime \TfitShort(\mud) = \Tsmud~ns.
The fitted amplitude of the long-lived component 
\AfitLong(\mud) = \Almud\%
is larger than zero with a significance of $3.8\sigma$.
Its fitted lifetime is 
\TfitLong(\mud) = \Tlmud~\mus. 
The physical values in Table~\ref{tab:amp_tau} include all further corrections.

%===============================================
%===============================================

\paragraph{{\bf Monte Carlo correction} ---}
The apparatus had been optimized for the Lamb shift
experiment~\cite{Pohl:2010:Nature,Antognini:2013:Science}.
For the present analysis of \TwoS{} lifetimes and populations, a
set of corrections has to be applied to the fit results, in order to gain
physical values.
As the transverse target dimensions are only $16 \times 25$~mm$^2$,
atoms in the \TwoS{} state may reach the target cell walls before RD
occurs.
Both \OneS{} and \TwoS{} atoms may also reach the walls after
x-ray emission, but before the muon decays.
The high-$Z$ target wall materials (Ag and ZnS) 
favor muon transfer and nuclear muon capture,
reducing the \mumi{} decay probability and thus the \xdele{} rate.
This leads to changes of the amplitudes and lifetimes of the observed
x-ray signals. 

A Monte Carlo simulation (MC) of the experiment was used to determine 
these loss correction factors of the observed signals.
The simulation included the 
experimental \Ekin{} distributions of 
\mup(\OneS) atoms~\cite{Kottmann:1999:Conf:EXAT98,Pohl:2001:PhD},
calculated cross sections for both \mup(\OneS) and \mup(\TwoS) 
scattering and \mup(\TwoS) quenching~\cite{Popov:2011:FormQuench2s, Popov:PC},
as well as the evolution of the atoms' \Ekin, 
the position-dependent x-ray detection efficiency, 
and the effect of the delayed electron time window ($t_e -t_X$).

For muonic deuterium 
we adapted \mup{} 
parameters with conservative systematic uncertainties.
We scaled the
initial \Ekin{}
by the
reduced mass ratio, 
a procedure justified due to the similarity of the cascade
in \mup{} and \mud~\cite{Lauss:1998:Xrays_muH,Lauss:1999:Cascade_muD}.
Since elastic scattering is unimportant at the low gas pressure
of 1~hPa, we also used the \mup{} cross sections in the \mud{} case.
According to the MC, detected \mup(\TwoS) atoms experience on average 
only 0.7 elastic collisions before RD
occurs. 
The \mup(\OneS) atoms undergo only 0.3 elastic
collisions before muon decay or arrival at the target walls.

The MC revealed that the fitted amplitudes \AfitShort{} have to be multiplied by 
\CAsmup{} (\mup) and  \CAsmud{} (\mud).
The fitted lifetimes of the fast components
\TfitShort{} are corrected by 
\CTsmup{} (\mup) and  \CTsmud{} (\mud),
because the fastest \TwoS{} atoms, which contribute most to the signal at early
times,
may reach the target wall before the beginning of the delayed electron time
window.

\begin{table}[t]
\begin{center}
  \caption{Results for \mup{} and \mud{} at 1~hPa.
    We give cascade times \TcascAvg, 
    total \TwoS{} populations \EpsTotal{} obtained from the x-ray yields, and 
    population \EpsShort{} and lifetime \TauShort{} of the radiatively 
    quenched short-lived \TwoS{} atoms.
    For the population of the long-lived \mud(\TwoS) atoms, see text.
    The populations, but not the lifetimes, have
    been corrected for $\mu$-decay.}
\label{tab:amp_tau}
\setlength{\tabcolsep}{0.04\columnwidth}
\begin{tabular}[c]{l | c c}
%\hline
%\multicolumn{3}{c}{Results:}\\
\hline \hline
                  & \mup{}              & \mud{} \\
\hline \hline
%\chisq{}/DOF      & \Chimup{}/\Dofmup{} & \Chimud{}/\Dofmud{}  \\
\TcascAvg{} (ns)  & \Tauavmup{}         & \Tauavmud{}   \\
\hline
\EpsTotal{} (\%)  & \PopTSmuH{} \cite{Anderhub:1984:KIntRatMup}
                  & \PopTSmuD{} \\
\hline
\EpsShort{} (\%)  & \AmpTwoSmup{}       & \AmpTwoSmud{} \\  
\TauShort{} (ns)  & \TauTwoSmup{}       & \TauTwoSmud{} \\
\hline
\EpsLong{} (\%)   & $1.10 \pm 0.08$ \cite{Pohl:2001:PhD} ~ ~
                                        & \AmpTwoSlmud{} (RD)
\footnote{to be multiplied by $(1/0.7 \times 8)$ (see Discussion)
\vspace{-3mm}} \\
\TauLong{} (\mus) & $1.04^{+0.29}_{-0.21}$~ \cite{Pohl:2006:MupLL2S} (CD)
                                        & \TauTwoSlmud{} (RD)\\
\hline\hline
\end{tabular} 
\end{center}
\vspace{-2mm}
\end{table}

\paragraph{{\bf X-ray yields} ---}
The lifetimes in Table~\ref{tab:amp_tau} are physical lifetimes, 
corrected as detailed above,
but \emph{not} for muon decay.
The physical \TwoS{} populations ($\varepsilon$) in 
Table~\ref{tab:amp_tau} have been multiplied
by $(1 - \TFit{}/\tau_\mu)$, where \TFit{} is the fitted lifetime,
to correct for muon decay 
($\tau_\mu = 2.2$~\mus), 
and by the \Ka{} yield \Yka{} to normalize to all muons.
In \mup, $\Yka(\mup) = \Ykamup$ at 1~hPa 
is interpolated from the values measured between
 0.33 and 8~hPa~\cite{Anderhub:1984:KIntRatMup}.

In \mud, yield measurements existed only for \Dtwo{} gas
pressures $\geq 17$~hPa~\cite{Bregant:1998:IntRatiosMup}.
We extract the \mud{} yields at 1~hPa from the fitted
amplitudes of the prompt \Ka, \Kb, and \Kr{} peaks determined here.
Using the \mup{} data
and the \mup{} yields
% interpolated from Ref.~\cite{Anderhub:1984:KIntRatMup}
allows one to determine the energy-dependent APD detection efficiency. This,
together with the prompt amplitudes fitted in \mud, gives
the \mud{} yields at 1~hPa of 
$\Yka = \Ykamud$,
$\Ykb = \Ykbmud$, and 
$\Ykr = \Ykrmud$.
As expected, the 
\mup{} and \mud{} x-ray yields are very similar~\cite{Bregant:1998:IntRatiosMup,Lauss:1998:Xrays_muH,Lauss:1999:Cascade_muD}.

\paragraph{{\bf Discussion} ---}
For muonic \emph{hydrogen} we find 
a cascade time of $\tcascH = \Tauavmup$~ns
and a radiatively quenched short-lived \TwoS{} 
population of $\EpsShort(\mup) = \AmpTwoSmup\%$
with lifetime $\TauShort(\mup) = \TauTwoSmup$~ns (Table~\ref{tab:amp_tau}).
These values (if gas pressure difference scaling is included) agree with, 
yet are more precise than, those in 
Ref.~\cite{Ludhova:2007:MupCascTime2S}.

The \mup{} data
are well fitted assuming no radiative quenching of long-lived \mup(\TwoS) atoms.
This agrees with previous observations~\cite{Pohl:2006:MupLL2S} that
long-lived \mup(\TwoS) atoms quench mainly via CD, an effect
explained both by direct CD~\cite{Popov:2011:FormQuench2s} and molecular
CD~\cite{Froelich:1993:Sidepath,Lindroth:2003:PPMU_with_erratum,
  Kilic:2004:Decay_ppmu}.  
Theory predicts that an excited \ppmu{} molecule decays mainly via CD, with
a radiative BR of only 2\% 
in muonic hydrogen~\cite{Lindroth:2003:PPMU_with_erratum,Kilic:2004:Decay_ppmu},
an effect too small to observe in the present experiment,
or in previous searches for RD
of long-lived 
\mup(\TwoS) atoms~\cite{Anderhub:1977:Search2S,Egan:1981:LongLived2S}.

The radiatively quenched short-lived \mup(\TwoS) atoms observed here,
$\EpsShort(\mup) = \AmpTwoSmup\%$,
and the  long-lived \mup(\TwoS) population from Ref.~\cite{Pohl:2001:PhD}, 
$\EpsLong(\mup) = 1.10 \pm 0.08$\%,
sum to $\EpsTotal(\mup) = 2.8^{+0.6}_{-0.4}$\%,
in agreement with the total \mup(\TwoS) population at 1~hPa,
$\EpsTotal(\mup) = \PopTSmuH\%$, 
calculated from the K-x-ray yields~\cite{Anderhub:1984:KIntRatMup}.

For muonic \emph{deuterium} we determine 
a cascade time of $\tcascD = \Tauavmud$~ns
and a radiatively quenched short-lived \TwoS{} 
population of $\EpsShort(\mud) = \AmpTwoSmud\%$
with lifetime $\TauShort(\mud) = \TauTwoSmud$~ns.
The short-lived amplitude agrees with the one observed in \mup.

Both the cascade time and lifetime of the short-lived \TwoS{} atoms 
scale as 
$\surd\overline{m(\mud)/m(\mup)} = 1.38$,
as expected for velocity-dependent collisional processes 
of \mup{} and \mud{} atoms with similar \Ekin.

The long-lived component observed in \mud{} is %the first
direct
evidence for RD
of long-lived (slow) \mud(\TwoS) atoms.
Its lifetime 
$\TauLong(\mud) = \TauTwoSlmud~\mus$ 
is in good agreement with the CD signal lifetime observed in \mup, 
$\TauLong(\mup) = 1.04^{+0.29}_{-0.21}~\mus$
 at 1~hPa~\cite{Pohl:2001:PhD,Pohl:2006:MupLL2S}.

The total \mud(\TwoS) population at 1~hPa is
$\EpsTotal(\mud) = \PopTSmuD\%$, calculated 
from the \mud{} yields measured here.
The difference $\EpsTotal - \EpsShort = \AmpLongExpected\%$ is the
\emph{expected} long-lived \mud(\TwoS) population.
It agrees with the value measured in \mup, 
$\EpsLong(\mup) = 1.10 \pm 0.08$\% \cite{Pohl:2001:PhD}.
Also, the size of the laser-induced $\TwoS \rightarrow \TwoP \rightarrow
\OneS$ signals (Fig.~\ref{fig:fit_time}) proves that the long-lived 2S
population is very similar in \mup{} and \mud.
The observed long-lived amplitude, $\EpsLong = \AmpTwoSlmud\%$ 
is $\sim 10$ times smaller than the expected amplitude.
This can be explained via molecular formation from the excited
\mud(\TwoS) state in a collision with \Dtwo,
%
% - - - - - - - - -
% Eq. (1)
%
\begin{equation}
  \label{eq:formation}
  \mud(\TwoS) + \Dtwo{} \rightarrow \ddmudee.
\end{equation}
Subsequent Auger-emission of both electrons 
leads to the formation of a
\ddmu{} molecular ion in a $^1S^e$ state of the $3d\sigma_g$ potential,
with vibrational quantum number 
$\nu = 2$ \cite{Pohl:2001:MolecQuenchMup,Lindroth:2003:PPMU_with_erratum,Kilic:2004:Decay_ppmu}.
This state decays with a radiative BR of $\sim 70\%$  
into the anti-binding $2p\sigma_u$ 
potential~\cite{Lindroth:2003:PPMU_with_erratum,Kilic:2004:Decay_ppmu}
%
% - - - - - - - - -
% Eq. (2)
%
\begin{equation}
  \label{eq:decay}
  \ddmu_{\nu=2} \rightarrow \mud(\OneS) + d + \gamma + \Ekin.
\end{equation}

The Franck-Condon principle predicts the 
x-ray spectrum in Fig.~5 of Ref.~\cite{Kilic:2004:Decay_ppmu}.
Moreover, the decay into the anti-binding $2p\sigma_u$ potential
produces \emph{accelerated} \mud(\OneS) atoms.
These can hit the walls before muon decay occurs which 
suppresses
the observed signal amplitude for the \xdele{} event class.
The MC predicts that acceleration to 15~eV will give
the observed factor of 1/7.

From the wave function~\cite{Kilic:2004:Decay_ppmu,Karr:PC} one finds that
27\%, 8\%, and 65\% of the \mud(\OneS) atoms formed by RD from 
the $\nu = 2$ state acquire a \Ekin{} of  
2.1, 16, and 56~eV, 
giving signal reduction factors of 0.34, 0.10, and 0.03, respectively.
This results in an overall signal reduction
% due to acceleration
 of 1/8,
in agreement with the observed factor of 1/7.

For muonic hydrogen, the absence of a long-lived radiative component in the
present data corresponds to a radiative BR that is at least 3.5 times smaller
(90\% C.L.) in \ppmu{} than the one observed in \ddmu.

The observed lifetime of the long-lived \TwoS{} component, 
$\TauLong{} \approx 1$~\mus{} at 1~hPa
both in \mup~\cite{Pohl:2006:MupLL2S} and \mud,
is given by
the excited muonic molecules formation rate 
(Eq.~(\ref{eq:formation}))~\cite{Wallenius:2001:mupDeExcMol}. 
Auger emission and RD/CD (Eq.~(\ref{eq:decay})) are much 
faster~\cite{Lindroth:2003:PPMU_with_erratum,Kilic:2004:Decay_ppmu}.

\paragraph{{\bf Conclusion} ---}

Comparison of 
cascade times, yields, lifetimes, and populations of the short- and 
long-lived \TwoS{} atoms presented here
shows that the cascade in muonic hydrogen and deuterium is 
very similar~\cite{Lauss:1998:Xrays_muH,Lauss:1999:Cascade_muD}.
A notable exception is the de-excitation of the long-lived \TwoS{} state, which
proceeds mainly via RD of \ddmu{} molecules in \mud,
in contrast to CD in \mup~\cite{Pohl:2006:MupLL2S}.

Molecular effects have proven to be important in muonic hydrogen 
scattering in gas targets~\cite{Adamczak:1996:Atlas,Pohl:2001:PhD},
as well as during muon-catalyzed fusion~\cite{Vesman:1967:muCF,Ponomarev:1990:MUCF,Balin:2011:MuCF_D2_HD}.
The present analysis demonstrates the prominence of molecular effects in
excited state processes of muonic hydrogen atoms. 
The measured formation rate of excited muonic molecules,
$\lambda \approx 0.5 \times 10^6~\mathrm{s}^{-1}$, at 1~hPa gas pressure is large, and
at gas pressures approaching liquid hydrogen density, as many as 65\% of the 
muons populate the metastable \TwoS{} 
state~\cite{Jensen:2002:Colldeexcite_Cascade}. 
Hence, formation and decay of excited muonic molecules from states with 
$n \geq 2$
is expected to have a strong influence on the cascade and dynamics of 
muonic hydrogen isotopes~\cite{Jensen:2002:Colldeexcite_Cascade,Froelich:1993:Sidepath}.

\paragraph{{\bf Appendix: Energy shift of the \Kr{} component} ---}

\begin{figure}[tb]
\begin{center}
\includegraphics[angle=0,width=0.9\columnwidth]{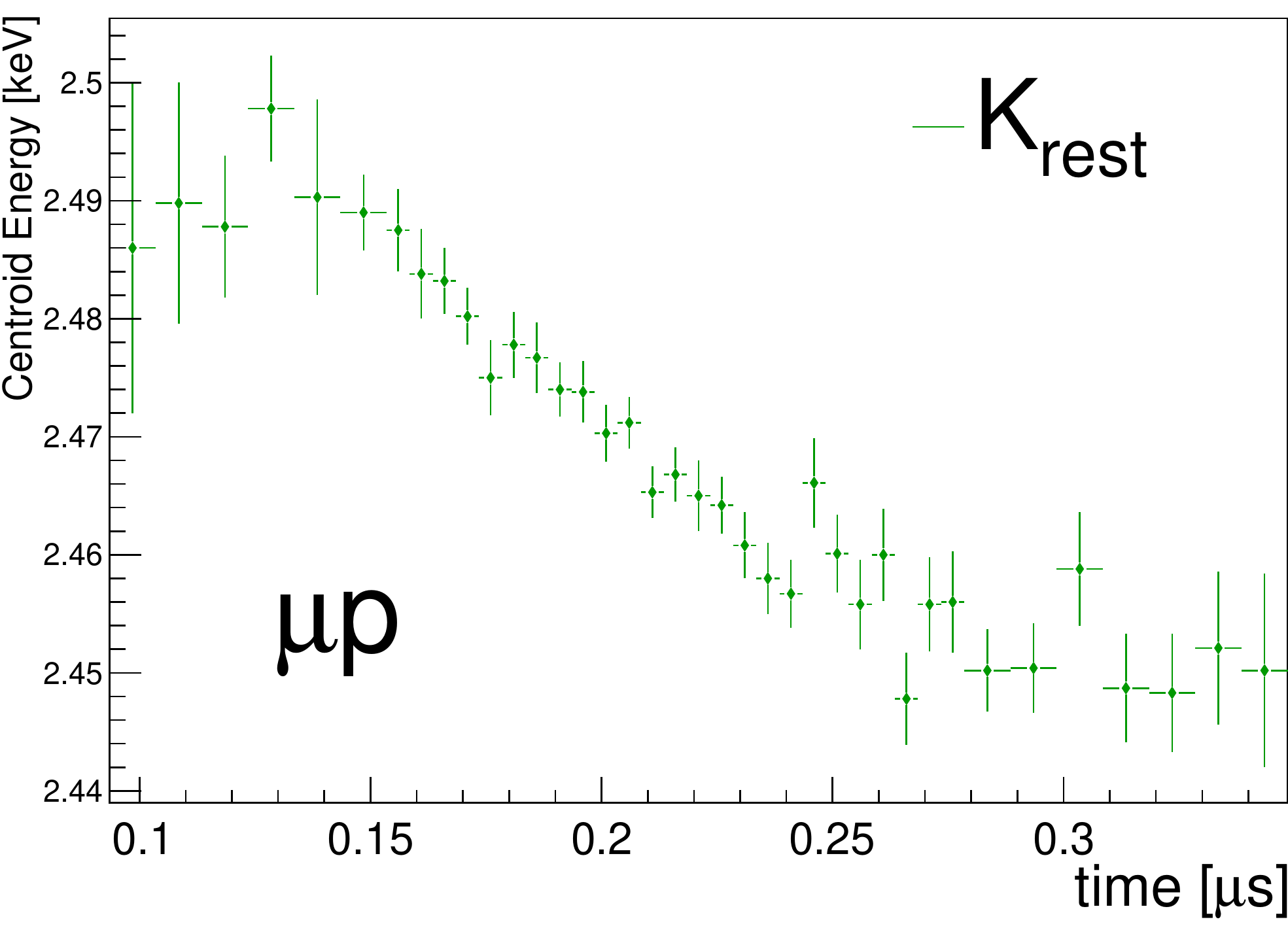}
\caption{ (Color online) Centroid energy of the \Kr{} component
as obtained by the fit of the \mup{} dataset between 0.1 and 0.4 $\mu$s after muon entry. 
}
\label{fig:fit_Krshift}
\end{center}
\vspace{-5mm}
\end{figure}

While fitting the \emph{time slices} of the \xdele{} energy spectra,
it became apparent that a fixed parameterization of the \Kr{} contribution
was not suitable to describe the data on a satisfactory level.
In order to improve the quality of the fit, the centroid energy of the
\Kr{}-peak was treated as a free parameter in the high statistics region
of the datasets.
As can be seen in Fig.~\ref{fig:fit_Krshift}, the energy of the \Kr{}
contribution is $2.49\pm0.01$~keV at times shortly after the
muon entry, but then declines to
$2.45\pm0.01$~keV for the latest cascade times in the case of \mup{}.
This can be explained by recollecting that muonic hydrogen atoms are created in
highly exited states \cite{LeonBethe:1962:mesonicH} and de-excite in the
cascade through different processes ~\cite{Jensen:2002:Colldeexcite_Xsect,Jensen:2002:Colldeexcite_Cascade}. 
At early time directly after muonic atom formation, \Kr{} transitions start
mainly from high n-levels (n $\sim$ 7-12) corresponding to transition energies
near the \mup{}(1S) dissociation energy of 2.53~keV. 
Lower initial n-levels (n $\sim$ 4-7) dominate at later times, leading to the
corresponding lower transition energies.

\paragraph{{\bf Acknowledgments} ---}
The authors thank V.P.~Popov and V.N.~Pomerantsev for the \mup(\OneS) and
\mup(\TwoS) cross sections used in the MC\, as well as J.P.~Karr and
L.~Hilico for the results of their calculation and fruitful discussions.
We acknowledge support from 
the European Research Council (ERC) under grant StG.~279765,
the Swiss National Science Foundation (projects 200020--100632 and 200021L--138175/1),
the Swiss Academy of Engineering Sciences,
the BQR de l'UFR de physique fondamentale et appliqu\'ee de l'Universit\'e 
Paris 6, 
the program PAI Germaine de Sta\"el no.~07819NH du minist\`ere des affaires 
\'etrang\`eres France,
and the FCT and FEDER under grant
SFRH/BPD/46611/2008 and project FIS/82006/2006.
P.I.~acknowledges support by the ``ExtreMe Matter Institute, Helmholtz Alliance
HA216/EMMI''.
T.W.H.~acknowledges support from the Max--Planck--Society and 
the Max--Planck--Foundation. Laboratoire Kastler Brossel is ``UMR n$^{\circ}$ 8552'' of the ENS, CNRS and UPMC.

\raggedright
%\bibliography{refs}
%

\end{document}